\documentclass[aps,pra,twocolumn,superscriptaddress]{revtex4-1}
\usepackage{color}
\usepackage{amsfonts}
\usepackage{amsmath}
\usepackage{amssymb}
\usepackage{graphicx}
\usepackage{float}
\usepackage{bbm}
\usepackage{bm}
\usepackage{tabularx}
\usepackage{dcolumn}
\usepackage{epstopdf}
\usepackage{epsfig}
\usepackage{verbatim}
\usepackage{array}
\usepackage[colorlinks,linkcolor=magenta,anchorcolor=blue,citecolor=blue,urlcolor=blue]{hyperref}

\usepackage{notes2bib}
\bibnotesetup{note-name = }
\usepackage[T1]{fontenc}
\usepackage{tgtermes}

\usepackage[normalem]{ulem}

\makeatletter
\@ifundefined{textcolor}{}
{%
 \definecolor{BLACK}{gray}{0}
 \definecolor{WHITE}{gray}{1}
 \definecolor{RED}{rgb}{1,0,0}
 \definecolor{GREEN}{rgb}{0,1,0}
 \definecolor{BLUE}{rgb}{0,0,1}
 \definecolor{CYAN}{cmyk}{1,0,0,0}
 \definecolor{MAGENTA}{cmyk}{0,1,0,0}
 \definecolor{YELLOW}{cmyk}{0,0,1,0}
}

\makeatother
\begin{document}

%%%%%%%%%%%%%%%%%%%%%%%%%%%%%%%
% MAIN \subfile{main.tex}
%%%%%%%%%%%%%%%%%%%%%%%%%%%%%%%
\widetext
\title{Mott insulating states of the anisotropic SU(4) Dirac fermions}
\author{Han Xu}
\affiliation{School of Physics and Technology, Wuhan University,
Wuhan 430072, China}
\affiliation{Department of Physics, City University of Hong Kong, Tat Chee Avenue, Kowloon, Hong Kong SAR, China}
\affiliation{City University of Hong Kong Shenzhen Research Institute, Shenzhen, Guangdong 518057, China}
\author{Yu Wang}
\email{yu.wang@whu.edu.cn}
\affiliation{School of Physics and Technology, Wuhan University,
Wuhan 430072, China}
\author{Zhichao Zhou}
\affiliation{School of Physics and Technology, Wuhan University,
Wuhan 430072, China}
\author{Congjun Wu}
\email{wucongjun@westlake.edu.cn}
\affiliation{New Cornerstone Science Laboratory, Department of Physics, School of Science, Westlake University, Hangzhou 310024, Zhejiang, China}
\affiliation{Institute for Theoretical Sciences, Westlake University, Hangzhou 310024, Zhejiang, China}
\affiliation{Key Laboratory for Quantum Materials of Zhejiang Province, School of Science, Westlake University, Hangzhou 310024, China}
\affiliation{Institute of Natural Sciences, Westlake Institute for Advanced Study, Hangzhou 310024, Zhejiang, China}

\begin{abstract}
We employ the large-scale quantum Monte-Carlo simulations
to investigate the Mott-insulating states of the half-filled SU(4) Hubbard model
on the square lattice with a staggered-flux pattern.
The noninteracting band structure that evolves from a
nested Fermi surface at zero flux to isotropic Dirac cones at
$\pi$ flux, exhibits anisotropic Dirac cones as the flux varies in between.
Our simulations show transitions between the three phases of Dirac semimetal,
antiferromagnet and valence-bond solid.
A direct continuous transition between the antiferromagnetic phase
and the valence-bond-solid phase is realized via varying the flux in the Mott regime. The simulated critical exponents remarkably agree with those
of SU(4) $J$-$Q$ model. Inside the valence-bond-solid phase induced by the flux, the plaquette valence-bond state with vanishing single-particle gap is identified.
At strong coupling, the valence-bond-solid phase disappears and the Mott-insulating state is always accompanied by antiferromagnetic ordering, regardless of the magnitude of the flux.
\end{abstract}

\maketitle

\section{Introduction}
%\textit{Introduction---}
Quantum phase transition has long been among the research foci of condensed matter physics, which is driven by quantum fluctuations in many-body systems \cite{Sondhi1997Continuous}.
Systems with enlarged degrees of freedom can exhibit SU($2N$) symmetry which can induce even richer states of matter \cite{Wu_2010,*Wu_2012,Cazalilla_2014}.
SU($2N$) symmetry is common in high energy physics, while it early emerged in condensed matter physics as a mathematical tool of the $1/N$-expansion for the purpose of handling strong correlation effects \cite{affleck1988largen,affleck1989largen,read1991largen,sachdev1991largen}.
After years of development, the SU(4) symmetry can arise in realistic solid-state systems: transition-metal oxides by tuning the parameters of the two-orbital model \cite{Li1998SU4,Pati1998Alternating,Joshi1999Elementary,Corboz2012Spin}, the $d^1$ systems such as $\ensuremath{\alpha}\text{-}{\text{ZrCl}}_{3}$ by combining spin and orbital to form the local moment of $j=3/2$ in the strong spin-orbit coupling limit \cite{Yamada2018Emergent}, and twisted bilayer systems where the layer pseudospin and the real spin are unified together \cite{Xu2018Topological,Zhang2021SU4}.
Beyond research on condensed matter, it has been proposed that large-spin ultra-cold fermions trapped in optical lattices provide highly tunable systems to explore exotic many-body SU($2N$) and Sp($2N$) physics \cite{Wu_PRL_2003,*Wu_MPLB_2006,Honerkamp_PRL_2004, Wu_2005,Hung_2011,Xu_2008,Chen_2005}.
Recent experiments have witnessed dramatic developments on SU($2N$) physics in cold atoms \cite{Taie2010Realization,zhang2014spectroscopic,Pagano2014,Hofrichter2016Direct,Riegger2018Localized,he2020collective,Song2020Evidence}, particularly the SU(6) Mott insulator \cite{Taie2012An} and antiferromagnetic correlations \cite{ozawa2018,Taie2022}.
The SU($2N$) fermionic atoms are more quantum-like, since the number of large spin components $2N$ enhances quantum fluctuations, while in contrast the SU(2) fermions with large composite spin magnitude $S$ are very classical-like. The fermionic atoms with SU($2N$) ($N>1$) symmetry can generate even richer quantum phases than electrons, which greatly enriches the understanding of quantum phase transitions.

The Hubbard model is a prototype model for studying Mott transitions of strongly correlated fermions.
The Mott-insulating states of the half-filled SU(2) model are associated with antiferromagnetic (AFM) order on a square lattice, independent of the flux \cite{Otsuka2002Mott,Chang2012,Toga2016,Parisen2015Fermionic,Otsuka2016}. In the half-filled SU(4) model, the minimal
SU($2N$) symmetry beyond SU(2), the Mott transition on a square lattice is accompanied with the AFM ordering which varies non-monotonically with increasing Hubbard $U$, first ascending to a peak value and then descending to a finite residual value \cite{Wang2014Competing,Wang2019Slater,Kim2019Dimensional}.
By contrast, in the half-filled SU(4) models on both the honeycomb and $\pi$-flux square lattices, which exhibit the Dirac spectrum, the valence-bond-solid (VBS) order emerges in the Mott-insulating state \cite{Zhichao2016Mott,Zhichao2018}.

In this paper, we explore Mott physics in the half-filled staggered-flux SU(4) Hubbard model on a square lattice. The synthetic flux ($0<\phi \leq \pi$) coupled to the hopping of fermions is treated as a control parameter of phase transition.
Our sign-problem-free projector determinant quantum Monte Carlo (QMC) simulations demonstrate a continuous AFM-VBS quantum phase transition via tuning the flux, and interestingly identify a plaquette valence-bond ordering region with finite spin gaps but vanishing single-particle gaps.

%-----------------------------------------------------------------------------
\section{Model}
\begin{figure}[tb]
\includegraphics[width=0.9\linewidth]{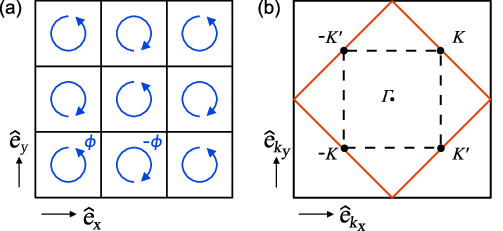}
\caption{
  (a) The square lattice with a staggered-flux pattern. The fermion hopping around a plaquette along the direction of the arrow acquires a phase $\phi$. (b) The Brillouin zone (red) and the Dirac points $ \mathbf{K}=(\frac{\pi}{2},\frac{\pi}{2})$, $ \mathbf{K}'=(\frac{\pi}{2},-\frac{\pi}{2})$.
  }\label{fig:nonint}
\end{figure}

%\textit{Model---}
We consider the half-filled SU(4) Hubbard model on the square lattice with a staggered-flux pattern,
\begin{equation}
  H=-\sum_{\langle ij\rangle,\alpha} \left( t_{ij}c_{i\alpha}^{\dagger}c_{j\alpha}+\mathrm{H.c.}\right)+\frac{U}{2}
\sum_i \left(n_i-2\right)^2,
\label{eq:hami}
\end{equation}
where $\langle ij\rangle$ denotes the nearest-neighbor (NN) sites and $\alpha$ is the flavor index running from 1 to 4.
$U$ describes the Hubbard repulsion, and $n_i=\sum_{\alpha=1}^4 c_{i\alpha}^{\dagger}c_{i\alpha}$ is the on-site particle number operator at site $i$.
The Hamiltonian satisfies the particle-hole symmetry, meaning that it is at half filling.
The NN hopping term is written as $t_{ij}=te^{i\theta_{ij}}$ where $t=1$  is set as the energy unit in our simulations.
As illustrated in Fig.~\ref{fig:nonint}(a), when hopping around a plaquette once,
the fermion acquires a phase $\sum_{\square}\theta_{ij}=(-1)^{i_x+i_y} \phi$
where the symbol $\sum_{\square}$ is the sum over the four sites in a plaquette and ($i_{x},i_{y}$) is the coordinate of the bottom-left site.
When $\phi=0,\pi$, the time-reversal symmetry is restored (let $|\theta_{ij}| =\phi/4$ for convenience). The staggered flux states were theoretically proposed in the study of strongly correlated systems such as high-$T_c$ cuprates \cite{affleck1988largen,Chakravarty2001Hidden}.
In cold atom experiment, the synthetic flux of the order of flux quantum per plaquette has been realized in optical lattices \cite{Dalibard2011Colloquium,aidelsburger2013realization,miyake2013realizing}. Moreover,
various tunable alternating flux patterns have been implemented on the square \cite{aidelsburger2011experimental}, triangular \cite{struck2013engineering} and honeycomb \cite{jotzu2014experimental} lattices by laser-assisted tunneling, lattice shaking and lattice modulation, respectively.

In the noninteracting limit, when $\phi=\pi$, there exist two low-energy Dirac cones located at $(\frac{\pi}{2},\pm\frac{\pi}{2})$ for each flavor and the Fermi velocity is $v_F=2t/\hbar$. When $0<\phi<\pi$, the Dirac cones become anisotropic, namely that the Fermi velocities around $(\frac{\pi}{2},\frac{\pi}{2})$ turn to be $\hbar v_F^{\perp}=2\sqrt 2 t\cos \frac{\phi}{4}$ along $(\hat{e}_{k_x}+\hat{e}_{k_y})/\sqrt{2}$ direction and $\hbar v_F^{\parallel}=2\sqrt 2 t\sin\frac{\phi}{4}$ along $(\hat{e}_{k_x}-\hat{e}_{k_y})/\sqrt{2}$ direction \cite{MySuppLink}.
%\bibnote[MySuppLink]{See Supplemental Material at URL, which includes Refs.~[], for details}.
With increasing $\phi$, the difference between $v_F^{\perp}$ and $v_F^{\parallel}$ decreases, leading to the suppression of the anisotropy of Dirac cones.

%-----------------------------------------------------------
%-----------------------------------------------------------

\section{AFM-VBS transition}
\begin{figure}[tb]
    \includegraphics[width=0.96\linewidth]{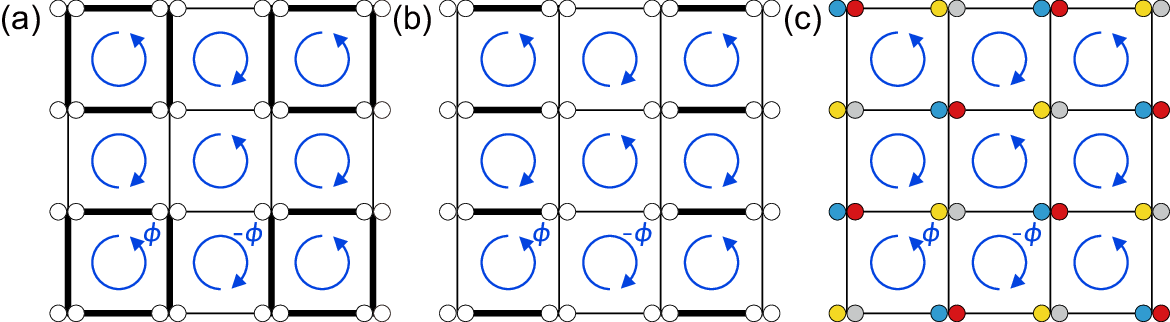}
    \caption{ Schematics of (a) pVBS, (b) cVBS and (c) AFM orders.
    }\label{fig:VBSpattern}
\end{figure}

% \textit{AFM-VBS transition---}
The projector determinant QMC method \cite{Assaad2008,Wu2005} is employed to investigate competing orders in our model via varying $\phi$ and $U$. At half-filling it is sign-problem-free and yields asymptotically exact results \cite{Wang2014Competing,Zhichao2018}.
Variation of flux modulates the anisotropy of Dirac cones, which in turn affects the orderings in the Mott-insulating states.
The AFM order takes place at the wavevector $\mathbf{Q}=(\pi,\pi)$.
The VBS orderings include the plaquette VBS (pVBS) and the columnar
VBS (cVBS) patterns as depicted in Figs.~\ref{fig:VBSpattern}(a) and \ref{fig:VBSpattern}(b), respectively.
The ordering wavevector for the cVBS is $\mathbf{Q}_{x}=(\pi,0)$, or,
$\mathbf{Q}_{y}=(0,\pi)$.
The pVBS is a superposition of cVBS orderings both at
$\mathbf{Q}_{x}$ and $\mathbf{Q}_{y}$.

%-----------------------------------------------------------
%-----------------------------------------------------------

The correlation functions and structure factors are used to characterize the AFM and VBS ordering.
For the AFM order, the equal-time SU($2N$) spin-spin correlation function is defined as $S(i,j)=\sum_{\alpha,\beta}\langle S_{\alpha\beta}(i)S_{\beta\alpha}(j)\rangle$
where $S_{\alpha\beta}(i)=c_{i,\alpha}^{\dagger}c_{i,\beta}
-\frac{\delta_{\alpha\beta}}{2N}\sum_{\gamma=1}^{2N}
c_{i,\gamma}^{\dagger}c_{i,\gamma}$ are the spin generators for a SU($2N$) group.
The spin structure factor is then defined via the spin-spin correlation functions,
\begin{equation}
\chi_{S}(\mathbf{Q})=\frac{1}{L^2}\sum_{ij}S(i,j)e^{i\mathbf{Q}\cdot\bm{r}_{ij}},
\end{equation}
where $\bm{r}_{ij}=\bm{r}_i-\bm{r}_j$ is the relative position vector between sites $i$ and $j$.
Then the AFM order parameter is given by the expression $\lim_{L\to\infty}\sqrt{\chi_S(\mathbf{Q})}/L$.
For the VBS order, the bond operator $d_{i,\hat e_a}$ is defined via the kinetic energy,
$d_{i,\hat e_a}=\sum_{\alpha=1}^{2N}\left(t_{i,i+\hat e_a}c_{i\alpha}^{\dagger}c_{i+\hat e_a,\alpha}+\mathrm{H.c.}\right)$
where $\hat e_{a}$ ($a=x, y$) are the primitive lattice vectors of the square lattice. The VBS structure factor is then defined in terms of the bond-bond correlations,
\begin{equation}\label{eq:defi_vbs}
  \chi_{D}=\sum_{a={x, y}}\chi_{D,a}(\mathbf{Q}_{a})=\sum_{a={x, y}}\frac{1}{L^2}\sum_{ij}\langle d_{i,\hat e_{a}}d_{j,\hat e_{a}}\rangle e^{i\mathbf{Q}_{a}\cdot\bm{r}_{ij}}.
\end{equation}
The expression $\lim_{L\to\infty}\sqrt{\chi_D}/L$ serves as the VBS order parameter.
Note that, due to the suppression of the overall kinetic energy scale, the VBS order parameter defined by Eq.~\eqref{eq:defi_vbs} decreases to the order of magnitude of $|t|^2/U$ in the large-$U$ regime \cite{Zhichao2016Mott,Zhichao2018}.

To locate the transition point more accurately,
we consider the AFM and VBS correlation ratios,
\begin{equation}
    \begin{aligned}
  R_{\mathrm{AFM}}&=1-\frac{1}{4}\sum_{a={x, y}}\frac{\chi_{S}(\mathbf{Q}
-\mathbf{\delta q}_{a})+\chi_{S}(\mathbf{Q}
+\mathbf{\delta q}_{a})}{\chi_{S}(\mathbf{Q})},\\
  R_{\mathrm{VBS}}&=1-\frac{1}{4}\sum_{a={x, y}}\sum_{b={x, y}}\frac{\chi_{D,a}(\mathbf{Q}_{a}+\mathbf{\delta q}_{b})}{\chi_{D,a}(\mathbf{Q}_{a})},
    \end{aligned}
\end{equation}
with $\mathbf{\delta q}_{x}=(\frac{2\pi}{L},0)$ and $\mathbf{\delta q}_{y}=(0,\frac{2\pi}{L})$ , which approach their size-independent values for sufficiently large $L$. Note that the ratio between the SU(4) AFM and VBS structure factors, $\chi_S(\mathbf{Q})/\chi_D$, is independent of the lattice size at the AFM-VBS critical point due to the emergence of enlarged symmetries \cite{Xiaoyan2019Monte}.

%-----------------------------------------------------------
\begin{figure}[tb]
  \includegraphics[width=0.9\linewidth]{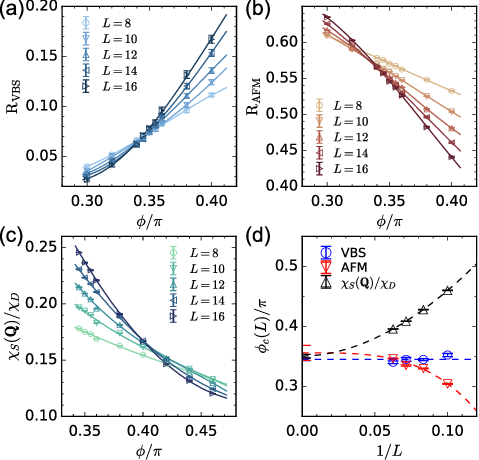}
  \caption{(a) The VBS correlation ratio; (b) the AFM correlation ratio; (c) the ratio between the SU(4) AFM and VBS structure factors;
  (d) the finite-size extrapolation of the crossing points. Here $U/t=10$.
  %The coupling strength is set to be $U/t=10$.
  }\label{fig:binder_u10}
\end{figure}
%-----------------------------------------------------------

%-----------------------------------------------------------
\begin{figure}[bt]
  \includegraphics[width=0.9\linewidth]{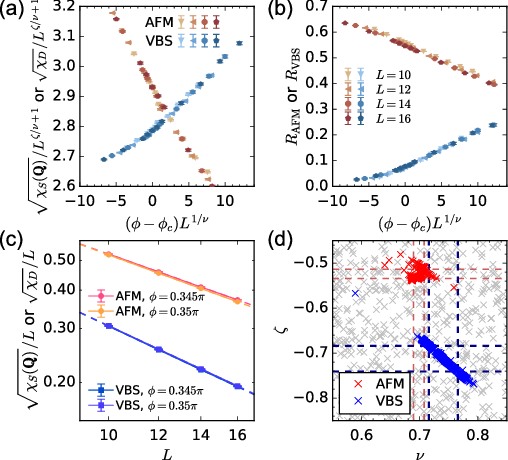}
  \caption{The scaling collapse of (a) structure factors and (b) correlation ratios at $U/t=10$. (c) The log-log plot of order parameters versus $L$ in the vicinity of critical point. (d) Best-fitting analysis of the critical exponents $\zeta$ and $\nu$ simultaneously. The converged values are colored while the initial guess values are greyed. The dashed lines represent the standard errors.
  % ($\chi_S(\bf Q)$ and $\chi_D$) ($R_{\rm AFM}$ and $R_{\rm VBS}$) $\sqrt{\chi_S(\bf Q)}/L$ and $\sqrt{\chi_D}/L$
  }\label{fig:fss}
\end{figure}
%-----------------------------------------------------------

We shall demonstrate the AFM-VBS Mott transition at $U/t=10$ by varying the flux $\phi$.
Figures~\ref{fig:binder_u10}(a), \ref{fig:binder_u10}(b) and \ref{fig:binder_u10}(c) show respectively plots of $R_{\mathrm{VBS}}$, $R_{\mathrm{AFM}}$ and $\chi_S(\mathbf{Q})/\chi_D$ as a function of $\phi$. In each case, the crossing points of curves for lattice sizes $L$ and $L-2$ can be fitted into the curve equation  $\phi_c(L)=\phi_c+aL^{-b}$, and then the critical point is obtained in the $L\to\infty$ limit \cite{Parisen2015Fermionic,shao2016quantum,Weber2018Two},  according to which the critical fluxes are found to be $\phi_c/\pi=0.345(3)$ for VBS, $\phi_c/\pi=0.356(13)$ for AFM and $\phi_c/\pi=0.350(3)$ for $\chi_S(\mathbf{Q})/\chi_D$, as shown in Fig.~\ref{fig:binder_u10}(d). The consistency of the critical fluxes infers a direct continuous AFM-VBS transition in our model.

Figures~\ref{fig:fss}(a) and \ref{fig:fss}(b) show the scaling collapse of the structure factors and correlation ratios, respectively.
In the vicinity of the critical point,
they obey the scaling law characterized by the anomalous scaling dimension $\eta$, the correlation length critical exponent $\nu$, and the dynamical critical exponent $z$ as
\cite{Parisen2015Fermionic,Weber2018Two,Gazit2018,Janke2008Monte},
\begin{eqnarray}
\chi(\delta\phi,L)&=&L^{-(d+z-4+\eta)} \tilde{\chi}\left(\delta\phi L^{\frac{1}{\nu}}
\right),
\\
R(\delta\phi,L)&=&\tilde{R} \left(\delta\phi L^{\frac{1}{\nu}}\right),
\end{eqnarray}
where $d=2$ is the spatial dimensionality; $\delta\phi=\phi-\phi_c$ is the deviation from the critical point $\phi_c$, and $\tilde{\chi}$ and $\tilde{R}$ are the scaling functions; $z=1$ \cite{senthil2004deconfined, Senthil2004Quantum, Sandvik2007,Wang2019Slater}.
We define $\zeta=-\nu(d+z-2+\eta)/2$ for convenience.
Typically, $\eta$ can be obtained from the slope of the log-log plot of $\chi$ versus $L$ at the critical
point $\phi_c/\pi=0.35$, as shown in Fig.~\ref{fig:fss}(c). $\eta$ is found to be $0.50$ and $0.92$ via the linear
fitting for the AFM and VBS orders, respectively.
Then, the best-fitting analysis is performed in Fig.~\ref{fig:fss}(d) where $\zeta$ and $\nu$ are obtained simultaneously
by optimizing the scaling collapse \cite{Houdayer2004Low,Melchert2009autoScale,*Sorge2015Pyfssa}.
The critical exponents are summarized as follows: $\nu_{\rm AFM}=0.70(1)$, $\eta_{\rm AFM}=0.50(4)$; and $\nu_{\rm VBS}=0.74(3)$, $\eta_{\rm VBS}=0.92(2)$.
The AFM and VBS orders give rise to similar values of exponent $\nu$ but different $\eta$ due to the simplified definition of VBS order parameter.
Our QMC results of $\nu$ and $\eta$  %in Eqs.~\eqref{eq:expo_nu} and \eqref{eq:expo_eta}
are almost identical to those obtained from the SU(4) $J$-$Q$ model \cite{Lou2009Anti,Kaul2011Quantum} where $\nu=0.70(2)$ and $\eta_{\rm AFM}=0.42(5)$, except for $\eta_{\rm VBS}=0.64(5)$ due to different definitions of the VBS order parameter. Therefore, the AFM-VBS transitions in our model and the SU(4) $J$-$Q$ model belong to the same universality class.

%-----------------------------------------------------------
\begin{figure}[tb]
  \includegraphics[width=0.9\linewidth]{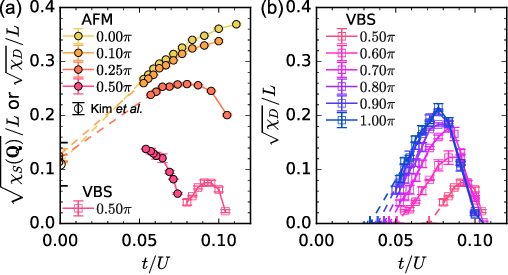}
  \caption{The finite-$U$ extrapolation of the order parameters. (a) The AFM (full circle) order parameters for $\phi/\pi\leqslant0.5$ and the VBS (open square) order parameter at $\phi/\pi=0.5$. (b) The VBS order parameters for $\phi/\pi\geqslant0.5$.
  }\label{fig:order_inf}
  \end{figure}
%-----------------------------------------------------------

\section{Large-$U$ Mott physics}
% \textit{Large-$U$ Mott physics---}
The AFM and VBS order parameters for various $U$ are obtained by finite-size extrapolation, and then extrapolated to the large-$U$ limit using a linear function of $t/U$.
As shown in Fig.~\ref{fig:order_inf}(a), the AFM order parameter for finite $U$ decreases with $\phi$, and in the large-$U$ limit the AFM order parameters for $\phi/\pi=0, 0.1$ and $0.25$ reasonably converge to the same extrapolated value $0.12(1)$ that is also found in the SU(4) Heisenberg model \cite{Kim2019Dimensional}. At $\phi/\pi=0.5$, the VBS-AFM transition occurs with increasing $U$. Figure~\ref{fig:order_inf}(b) shows that the VBS order parameters vanish for $U>30$, regardless of $\phi$.

\begin{figure}[tbh]
    \includegraphics[width=0.96\linewidth]{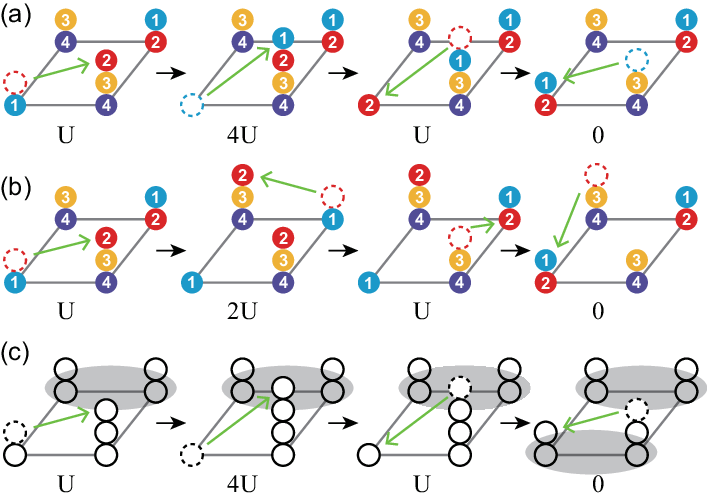}
    \caption{The fourth-order perturbation theory for the energy scales of AFM and VBS orderings. (a) The superexchange process occurring exclusively between two sites.
    (b) The ring-exchange process connecting four lattice sites, relevant to AFM ordering.
    (c) The superexchange process corresponding to the fourth-order perturbation term in a two-site singlet bond. Interaction energies are respectively indicated at the bottom of each configuration.
    }\label{fig:ring-exchange}
\end{figure}

The suppression of AFM ordering with increasing flux $\phi$ can be attributed to the ring exchange process, which can be understood by considering the N\'eel and SU($2N$) bond singlet configurations on a plaquette of the square lattice, as depicted in Fig.~\ref{fig:ring-exchange}.

To maintain the N\'eel configuration, spin-flip processes are not allowed.
In Fig.~\ref{fig:ring-exchange}(a), two fermions of certain species on one site can hop to the NN sites, but they must hop back, cancelling out the staggered phase $\pm\phi$.
The number of NN sites is denoted by the coordination number $c$, and then the total number of the virtual hopping processes is $c N(N-1)$.
The interaction energy due to the virtual hopping process is indicated for each configuration in Fig.~\ref{fig:ring-exchange}.
Therefore, the energy gain at the fourth-order superexchange level is $c N(N-1)\frac{t^{4}}{4U^3}\to c N^2J'$ with $J'=\frac{t^4}{U^3}$.
Other types of fourth-order superexchange processes \cite{Zhichao2018} that own the same energy scale of $N^2$ are not depicted here.

Furthermore, as illustrated in Fig.~\ref{fig:ring-exchange}(b), one fermion of a certain species undergoes a (counterclockwise) ring-exchange process around the plaquette, acquiring a staggered phase factor $e^{\pm i\phi}$. There is also the (clockwise) ring-exchange process with a staggered phase factor $e^{\mp i\phi}$ around the same plaquette. The total number of the virtual hopping processes is $c N$. Hence, the energy gain of the fourth-order ring-exchange process is $c N\frac{t^{4}}{U^3}\cos{\phi}\to c NJ_{\square}$ with $J_{\square}=\frac{t^4}{U^3}\cos{\phi}$.

In contrast, spin-flip processes are allowed in the SU($2N$) kinetic dimer and one site can only involve in the formation of a single singlet, as depicted in Fig.~\ref{fig:ring-exchange}(c).
The energy scale of the fourth-order superexchange process is $N(N-1)(N+2)(N+1)\frac{t^4}{4U^3}\to N^4J'$.

Similarly, the energy gains at the second-order perturbation level are $c NJ$ and $N(N+1)J\to N^2J$ for the N\'eel and bond singlet configurations, respectively, where $J=\frac{t^2}{U}$.

Overall, the energy scale of the AFM state is estimated at $c NJ + c N^2J' + c NJ_{\square}$, while the energy scale of the VBS state is estimated at $N^2J + N^4J'$.
Hence, increasing $\phi$ suppresses AFM due to the decrease of the ring-exchange term $J_{\square}$ at finite $U$, whereas for sufficiently large $U$, the ring-exchange process is inhibited so that the flux stops affecting AFM ordering.
In contrast, the VBS state is dimerized in which case the ring exchange process does not contribute at the leading level, and thus is not affected by the increasing flux.

%-----------------------------------------------------------
%-----------------------------------------------------------

\section{Gap opening mechanism}

%-----------------------------------------------------------
\begin{figure}[tb]
\includegraphics[width=0.85\linewidth]{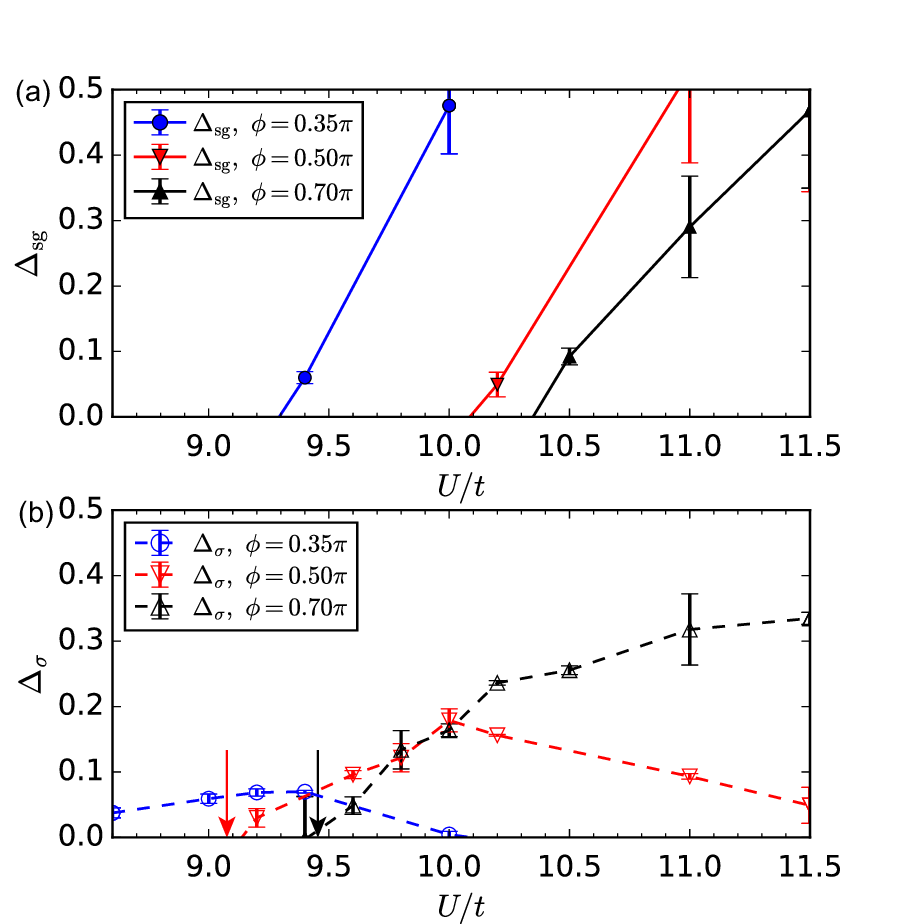}
  \caption{In the $L\to\infty$ limit, (a) the single-particle gap $\Delta_{\rm sg}$ and (b) the spin gap $\Delta_{\sigma}$ behave as a function of $U$ for various $\phi$. The arrows indicate the critical coupling strength determined by the VBS correlation ratio.
  }\label{fig:gap_inf}
\end{figure}
%-----------------------------------------------------------

The single-particle gap $\Delta_{\rm sg}$ and the spin gap $\Delta_{\sigma}$ are defined at the wave vector $\mathbf{K}=(\frac{\pi}{2},\frac{\pi}{2})$ and the AFM wave vector $\mathbf{Q}=(\pi, \pi)$, respectively.
In the weak-coupling regime, for $\phi>0$ the system lies in the semimetal phase. With the increase of $U$, the system undergoes Mott transition accompanied by single-particle gap opening. In the Mott-insulating regime, the spin gap opens in the VBS phase, but closes in the AFM phase.

In Figs.~\ref{fig:gap_inf}(a) and \ref{fig:gap_inf}(b), for various $\phi$, the $L\to\infty$ limits of $\Delta_{\rm sg}$ and $\Delta_{\sigma}$
are plotted as a function of $U$, respectively.
For $\phi/\pi=0.35$, the spin gap closes at $U/t\gtrsim10$, corresponding to the Goldstone modes in the AFM phase.
For $\phi/\pi=0.5$ and $0.7$, the spin gap opens at $U_c/t\approx9.1$ and $9.5$ respectively, which coincides with the emergence of VBS orders. This means that spin-gap opening and symmetry breaking occur simultaneously. Remarkably, in the coupling regime $9\lesssim U/t\lesssim 10$, $\Delta_{\rm sg}=0$ but $\Delta_{\sigma}>0$. This can be interpreted as a pVBS state, as explained below.

The kinetic bond operator between site $i$ and site $i+\hat e_a$ is defined as
$d_{i,\hat e_a}=\sum_{\alpha=1}^{2N}\big(t_{i,i+\hat e_a} c_{i\alpha}^{\dagger} c_{i+\hat e_a,\alpha}+\mathrm{H.c.}\big)$
where $a=x,y$.
Now, consider the bond-bond interaction term $-\frac{g}{2}\sum_{i,a} d_{i,\hat e_a} d_{i,\hat e_a}$, and combine it with the noninteracting Hamiltonian,
\begin{equation}
  H_0=-\sum_{\langle ij\rangle,\alpha} \big( t_{ij}c_{i\alpha}^{\dagger}c_{j\alpha}+\mathrm{H.c.}\big),
\end{equation}
we obtain
\begin{equation}
    H= H_0 - \frac{g}{2}\sum_{i,a} d_{i,\hat e_a} d_{i,\hat e_a}.
\end{equation}
The second term on the right-hand side can be expanded as
\begin{equation}
    \begin{aligned}
        d_{i,\hat e_a} d_{i,\hat e_a} = \sum_{\alpha\beta} \big(t_{i,i+\hat e_a}^2 c_{i\alpha}^{\dagger} c_{i\beta}^{\dagger}c_{i+\hat e_a,\beta} c_{i+\hat e_a,\alpha} \\
        +c_{i\alpha}^{\dagger} c_{i+\hat e_a,\alpha} c_{i+\hat e_a,\beta}^{\dagger}c_{i\beta} + \mathrm{H.c.}\big).
    \end{aligned}
\end{equation}
Since there is no long-range pairing order in our model, we can omit the first pair-hopping term when performing the VBS mean-field calculations.
Therefore, the problem now is to find the mean-field solution of the Hamiltonian
\begin{equation}
    H = H_0 - g\sum_{i,a}\sum_{\alpha\beta} c_{i\alpha}^{\dagger} c_{i+\hat e_a,\alpha} c_{i+\hat e_a,\beta}^{\dagger}c_{i\beta}.
\end{equation}
This allows us to construct the valence bond operator $\hat \chi_{i,\hat e_a}=g\sum_{\alpha}c_{i\alpha}^{\dagger} c_{i+\hat e_a,\alpha}$ \cite{affleck1988largen,affleck1989largen}.
By introducing the static Hubbard-Stratonovich field $\chi_{i,\hat e_a}$, the mean-field Hamiltonian is written as \cite{affleck1988largen,affleck1989largen}
\begin{equation}
        H_{\mathrm{MF}}=H_0 -\sum_{i,a,\alpha}\big(\chi_{i,\hat e_a}c_{i+\hat e_a,\alpha}^{\dagger}c_{i\alpha}+\mathrm{H.c.}\big),
    \label{eq:supp:mf_h}
\end{equation}
where $\chi_{i,\hat e_a} = g\sum_{\alpha}\langle c_{i\alpha}^{\dagger}c_{i+\hat e_a,\alpha}\rangle$.
For the VBS order, the mean-field valence bond order parameter is defined as
\begin{eqnarray}
    \chi_{i,\hat e_a}&=&\chi_a e^{-i \mathbf{Q}_a \cdot \bm{r}_i}, \label{eq:supp:x_vbs}\\
    \chi_{a}&=&g\sum_{\alpha}e^{i\mathbf{Q}_a\cdot \bm{r}_i}\langle c_{i\alpha}^{\dagger}c_{i+\hat e_{a},\alpha}\rangle,
\end{eqnarray}
where $e^{-i \mathbf{Q}_a \cdot \bm{r}_i}$ represents the periodic bonding strength along the $a$ axis. For convenience, the flavor index $\alpha$ is omitted hereafter.

Substituting Eq.~\eqref{eq:supp:x_vbs} into Eq.~\eqref{eq:supp:mf_h}, the mean-field Hamiltonian in the reciprocal space takes the form
\begin{equation}
    \begin{aligned}
    &H_{\mathrm{MF}}=-\sum_{\bm{k}}\big[a(\bm{k}) c_{A\bm{k}}^{\dagger} c_{B\bm{k}}+\mathrm{H.c.}\big]
      +\sum_{\bm{k}}\big[i\chi_x\sin{k_x} \\ &\big(c_{A,\bm{k}+\mathbf{Q}_x}^{\dagger}c_{B\bm{k}} + c_{B,\bm{k}+\mathbf{Q}_x}^{\dagger}c_{A\bm{k}}\big) + i\chi_y\sin{k_y} \big(c_{A,\bm{k}+\mathbf{Q}_y}^{\dagger}c_{B\bm{k}} \\ &+ c_{B,\bm{k}+\mathbf{Q}_y}^{\dagger}c_{A\bm{k}}\big)
      + \mathrm{H.c.}\big],
    \end{aligned}
    \label{eq:supp:mf_vbs}
\end{equation}
where $c_{\bm{k}}\equiv(c_{A\bm{k}},c_{B\bm{k}})$ is the basis of second quantized operators for sublattices $A$ and $B$, and $a(\bm{k})=e^{-i\phi/4}\cos{k_x}+e^{i\phi/4}\cos{k_y}$.
 In the basis of $(c_{\bf K}, c_{\bf -K'}, c_{\bf -K}, c_{\bf K'})$, the four zero-energy single particle states at Dirac points $|\pm \mathbf{K}\rangle$ and $|\pm \mathbf{K}'\rangle$ are connected by the valence bond orderings.
The effective Hamiltonian matrix is expressed as
\begin{equation}
  \begin{aligned}
  H_{\mathrm{MF}} &= -i\chi_x|\mathbf{K}\rangle\langle-\mathbf{K}'| - i\chi_y|\mathbf{K}\rangle\langle\mathbf{K}'| \\
  &+i\chi_x|-\mathbf{K}\rangle\langle\mathbf{K}'| + i\chi_y|-\mathbf{K}\rangle\langle-\mathbf{K}'| + \mathrm{H.c.}
  \end{aligned}
\end{equation}
By the degenerate perturbation theory, the first-order correction $\lambda$ is determined by
\begin{equation}\label{eq:supp:eqpvbs}
    \lambda^4-2\lambda^2(\chi_x^2+\chi_y^2)+(\chi_x^2-\chi_y^2)^2 = 0.
\end{equation}
For pVBS ordering (Fig.~\ref{fig:VBSpattern}(a)), $\chi_x=\chi_y$ and then a two-fold degeneracy remains (vanishing of single-particle gap).
By contrast, for cVBS ordering (Fig.~\ref{fig:VBSpattern}(b)), either $\chi_y=0$ or $\chi_x=0$, which completely lifts the degeneracy (gap opening). Thus the VBS state with zero single-particle gap is a pVBS.

%-----------------------------------------------------------
\begin{figure}[tb]
  \includegraphics[width=0.85\linewidth]{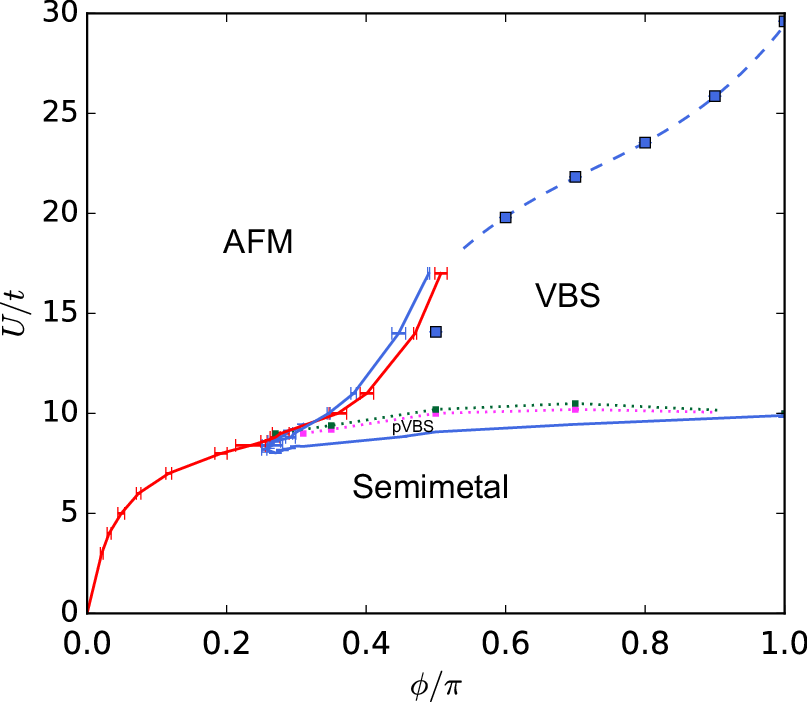}
  \caption{The phase diagram of the SU(4) Hubbard model on the square lattice with a staggered-flux pattern. Error bar on the semimetal-pVBS phase boundary is smaller than the line width.
  }\label{fig:phase_diagram}
\end{figure}
%-----------------------------------------------------------

\section{Phase diagram}
% \textit{Phase diagram---}
The phase diagram for our staggered-flux SU(4) Hubbard model is shown in Fig.~\ref{fig:phase_diagram}. The blue and red solid curves are respectively determined by the correlation ratios, which depict the phase boundaries between semimetal, AFM and VBS phases. The pink dotted curve is the ``boundary'' of pVBS phase with vanishing single-particle gap, while the green dotted curve is the ``boundary'' of VBS phase with single-particle gap opening (With denser data, the two ``boundaries'' should merge into one).
The blue dashed curve fits the blue-square points corresponding to the vanishing VBS order parameter. When $U/t\gtrsim11$, the blue and red solid curves between AFM and VBS phases do not precisely coincide owing to strong finite-size effects.

In the large-$U$ regime $U/t>30$, the Mott-insulating states always accompany with AFM ordering which is consistent with the SU(4) Heisenberg model. When $8.2\lesssim U/t\lesssim30$, the continuous AFM-VBS transitions occur at critical coupling $U_c(\phi)$ which increases with the flux $\phi$. Remarkably when $8.2\lesssim U/t \lesssim 9$, the $\phi$-induced AFM-pVBS transitions accompany with single-particle gap closing and spin gap opening. For $\phi/\pi\lesssim 0.25$, the $U$-induced Mott transitions occur with the emergence of AFM order, whereas for $0.25\lesssim\phi/\pi<1$ semimetal-pVBS transitions do not open a Mott gap.

%%%%%%%%%%%%%%%%%%%%%%%%%%%%%%%%%%%%%%%%%%%%%%%%%%%%%%%%%
% \textit{Summary---}
\section{Conclusions}
We have performed the large-scale QMC simulations to investigate the Mott-insulating states of the half-filled SU(4) Hubbard model on the square lattice with a staggered-flux pattern.
Increasing flux $\phi$ suppresses the single-particle gap opening and the AFM ordering, whereas the effects of the flux is completely inhibited in the large-$U$ limit where our model reduces to SU(4) Heisenberg model. With the influence of flux $\phi$, VBS phase emerges in the SU(4) Dirac fermions, and particularly the pVBS phase with zero single-particle gap is identified.
We have realized the continuous AFM-VBS transition in the SU(4) Hubbard model via tuning the flux $\phi $ when $8.2\lesssim U/t\lesssim30$.
The critical exponents calculated by QMC simulations remarkably agree with those of SU(4) $J$-$Q$ model.
In comparison with other SU(4) models for AFM-VBS transitions, our model opens a new avenue for exploring the AFM-VBS transition in a prototype model (e.g. SU(4) Hubbard model) by tuning a realistic controllable parameter (e.g. synthetic flux $\phi$ in optical lattices).
In general the emergence of VBS order accompanies with opening of single-particle gap and spin gap. In our simulations we have found a pVBS region of $8.2\lesssim U/t \lesssim 10$ and $0.25\lesssim\phi/\pi<1$ where single-particle gap closes. This result extends the understanding of VBS states.

It is worth noting that spin gap calculations using imaginary-time Green's function formalism may not distinguish a "pseudo gap" due to overdamped spin excitations around ($\pi, \pi$), which is left for future study.

\section*{Acknowledgments}
% \textit{Acknowledgments---}
This work is financially supported by
the National Natural Science
Foundation of China under Grants No. 11874292,
No. 11729402, and No. 11574238.
C.W. is supported by the New Cornerstone Science Foundation and the National Natural Science Foundation of China
under the Grants No. 12174317 and No. 12234016. We
acknowledge the support of the Supercomputing Center of
Wuhan University.

%%%%%%%%%%%%%%%%%%%%%%%%%%
    % Bib Reference
%\bibliography{../stagger_ref}

%%%%%%%%%%%%%%%%%%%%%%%%%%

\end{document}